\documentclass[aps,prl,reprint,superscriptaddress]{revtex4-1}

\usepackage{graphicx}% Include figure files
\usepackage{dcolumn}% Align table columns on decimal point
\usepackage{bm}% bold math

\begin{document}

% Use the \preprint command to place your local institutional report
% number in the upper righthand corner of the title page in preprint mode.
% Multiple \preprint commands are allowed.
% Use the 'preprintnumbers' class option to override journal defaults
% to display numbers if necessary
%\preprint{}

%Title of paper
\title{Constraints on New Gravitylike Forces in the Nanometer Range}

% repeat the \author .. \affiliation  etc. as needed
% \email, \thanks, \homepage, \altaffiliation all apply to the current
% author. Explanatory text should go in the []'s, actual e-mail
% address or url should go in the {}'s for \email and \homepage.
% Please use the appropriate macro foreach each type of information

% \affiliation command applies to all authors since the last
% \affiliation command. The \affiliation command should follow the
% other information
% \affiliation can be followed by \email, \homepage, \thanks as well.
\author{Y. Kamiya}
\email[]{kamiya@icepp.s.u-tokyo.ac.jp}
%\homepage[]{Your web page}
%\thanks{}
%\altaffiliation{}
\affiliation{International Center for Elementary Particle Physics and Graduate School of Science, the University of Tokyo, Tokyo 113-0033, Japan}

\author{K. Itagaki}
\thanks{Deceased.}
\affiliation{International Center for Elementary Particle Physics and Graduate School of Science, the University of Tokyo, Tokyo 113-0033, Japan}

\author{M. Tani}
\affiliation{International Center for Elementary Particle Physics and Graduate School of Science, the University of Tokyo, Tokyo 113-0033, Japan}

\author{G. N. Kim}
\affiliation{Department of Physics, Kyungpook National University, Daegu 702-701, Republic of Korea}

\author{S. Komamiya}
\affiliation{International Center for Elementary Particle Physics and Graduate School of Science, the University of Tokyo, Tokyo 113-0033, Japan}

%Collaboration name if desired (requires use of superscriptaddress
%option in \documentclass). \noaffiliation is required (may also be
%used with the \author command).
%\collaboration can be followed by \email, \homepage, \thanks as well.
%\collaboration{}
%\noaffiliation

\date{\today}

\begin{abstract}
We report on a new constraint on gravitylike short-range forces,
in which the interaction charge is mass,
obtained by measuring the angular distribution of 5 \AA\ neutrons scattering off atomic xenon gas.
Around $10^7$ scattering events were collected 
at the 40 m small angle neutron scattering beam line
located at the HANARO research reactor of the Korean Atomic Energy Research Institute.
The extracted coupling strengths of new forces in the Yukawa-type parametrization
are $\hat{g}^2 = (0.2 \pm 6.8 \pm 2.0) \times 10^{-15}$ GeV$^{-2}$ 
and $\hat{g}^2 = (-5.3 \pm 9.0 {\,}^{+ 2.7}_{-2.8})  \times 10^{-17}$ GeV$^{-2}$
for interaction ranges of 0.1 and 1.0 nm, respectively.
These strengths correspond to 
95\% confidence level limits of
$g^2 < (1.4 \pm 0.2) \times 10^{-14}$ GeV$^{-2}$ 
and $g^2 < (1.3 \pm 0.2) \times 10^{-16}$ GeV$^{-2}$,
improving the current limits for interaction ranges between 4 and 0.04 nm
by a factor of up to 10.
\end{abstract}

% insert suggested PACS numbers in braces on next line
% insert suggested keywords - APS authors don't need to do this
%\keywords{}

%\maketitle must follow title, authors, abstract, \pacs, and \keywords
\maketitle

% body of paper here - Use proper section commands
% References should be done using the \cite, \ref, and \label commands
Extensions of the standard model of particle physics have long been discussed. 
Some of these theories, based on supersymmetry or extra space dimensions, 
naturally involve gravity or gravitylike interactions even at low energies.
Several models predict new bosons mediating gravitylike forces that couple to mass, baryon number, 
or, in the case of grand unification models, to the difference of baryon and lepton numbers 
\cite{Fayet:1993fu, Fayet:2001ve, ArkaniHamed:2002kj, Dimopoulos:2003hb, Arkani-Hamed:359727, Randall:389728}.
This class of models induces modifications to the Newtonian inverse-square law
of gravitational interactions, and may also cause violation of the weak equivalence principle. 
These proposals motivate us to search for new gravitylike forces.
Comprehensive reviews of 
such theories, and the forces they predict, can be found in Refs.
\cite{Kim:1986ax, Adelberger:2003ik, Adelberger:2009gg}.

\begin{figure}
\includegraphics[width=8.2cm]{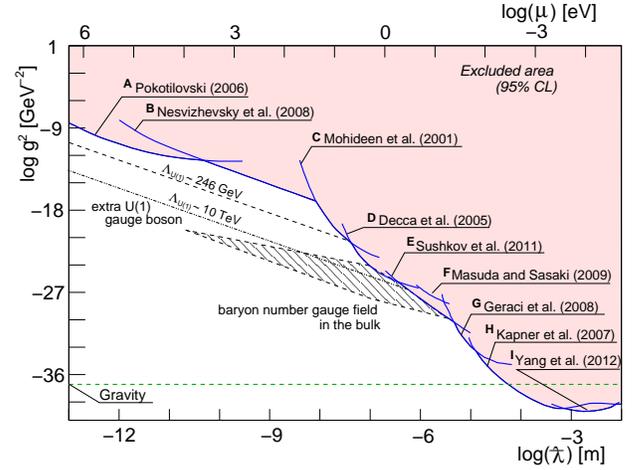}%
\caption{\label{fig1}
Experimental limits at 95\% C.L. for new gravitylike 
forces in the Yukawa-type parametrization space.
Constraints A and B were obtained by measuring neutron-atom scattering \cite{Pokotilovski:2006gd, Nesvizhevsky:2008jz},
while constraints C to I were obtained using macroscopic methods
\cite{BORDAG:2001tt, Decca:2005hf, Sushkov:2011bq, Masuda:2009gk, Geraci:2008hp, Kapner:2007hh, Yang:2012dy}.
Theoretical expectations from an extra $U(1)$ gauge boson at different symmetry breaking scales 
$\Lambda_{U(1)}$ ($\sim$ 246 GeV and 10 TeV) \cite{Fayet:1993fu, Fayet:2001ve},
and from a baryon number gauge field in the bulk of extra space dimensions \cite{ArkaniHamed:2002kj, Dimopoulos:2003hb}
are shown as dashed lines and the hatched area, respectively.
}
\end{figure}

The forces due to such new bosons can be simply modeled by 
a Yukawa-type scattering potential written in natural units as
\begin{equation}
V_{new}(r) = - \frac{1}{4\pi} g^2 Q_1 Q_2 \frac{e^{-\mu r}}{r}~,
\end{equation}
where $g^2$ is a coupling strength, $Q_i$ are coupling charges,
and $\mu$ is the mass of the boson mediating the force.
Such models can be considered in the 
$g^{2}-\mu$ or $g^{2}-\lambdabar$ 
parameter space,
where $\lambdabar \equiv 1/\mu$ is the interaction range.
Current experimental limits at 95\% confidence level (C.L.)
for interactions that couple to mass are shown in Fig.\,\ref{fig1}.
Constraints A and B were obtained by microscopic experiments that 
precisely measured interactions between neutrons and atoms \cite{Pokotilovski:2006gd, Nesvizhevsky:2008jz}. 
The experimental method used to achieve the results
reported in this Letter follows a similar approach.
Constraints C to I were obtained by macroscopic tests,
searching for non-Newtonian forces between test masses
using techniques such as torsion balances and microcantilevers
\cite{BORDAG:2001tt, Decca:2005hf, Sushkov:2011bq, Masuda:2009gk, Geraci:2008hp, Kapner:2007hh, Yang:2012dy}.
%=== fig 1 =====
%=== fig 1 =====

\begin{figure}
\includegraphics[width=8.2cm]{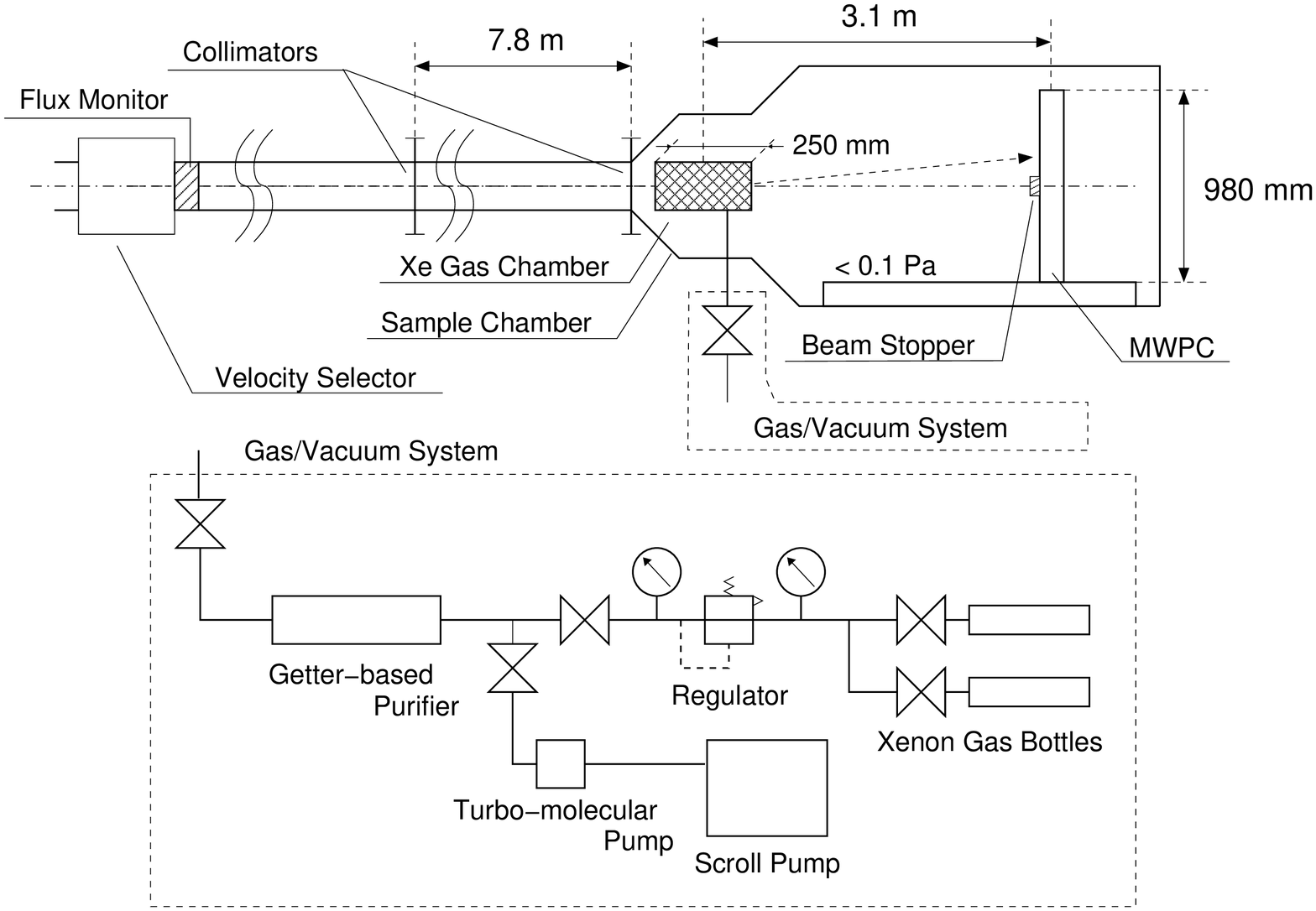}%
\caption{\label{fig2}
A schematic view of the experimental setup (not to scale).
Neutrons of wavelength 5 \AA\ and a 12\% spread were chosen by a velocity selector.
Two collimators of diameter 22 mm produced a neutron beam
with 3 mrad. divergence, which impinged on a xenon-filled target chamber of length 250 mm. 
Scattered neutrons were measured by a MWPC with pixel size 5.1 $\times$ 5.1 mm$^2$ located 3.11 m downstream of the gas target.
The xenon gas was purified to better than 10 ppm by a getter system
using a carefully controlled flow rate.
}
\end{figure}

This experiment was performed at the 40 m small angle neutron scattering (SANS) beam line
located at the HANARO research reactor of the Korean Atomic Energy Research Institute \cite{Han:2013gg}.
Figure\,\ref{fig2} shows a schematic drawing of the experimental apparatus.
The angular distribution of neutrons scattered by a xenon gas target was measured, and deviations from
the expectations from known interactions were used to set limits on additional, unknown, interactions.

Neutrons with an average wavelength of 5 \AA\ and a spread of 12\% (FWHM) were used,
selected by appropriately setting the rotation rate and tilt angle of a helical slot velocity selector 
installed at the entrance to the beam line.
The neutron beam intensity was monitored 
by a ${}^3$He--filled sampling gas counter
located immediately downstream of the velocity selector.
Two collimators made of sintered B${}_4$C with a circular aperture of diameter 22 mm were placed
12.2 m and 20.0 m downstream of the gas counter, defining the beam divergence of 3 mrad.
%=== fig 2 =====
%=== fig 2 =====

A cylindrical target chamber with an inner volume of length 250 mm and diameter 130 mm
was placed in the sample chamber immediately downstream of the second collimator.
The target chamber was connected to a getter-based purifier which reduced contaminants in the xenon gas to less than 10 ppm,
a vacuum system consisting of turbomolecular and scroll pumps, and xenon gas bottles.
The neutron entrance and exit windows of the target chamber were made of a single Si crystal \cite{1976NucIM.135..289B},
and only metals were used in the vacuum seals to avoid contamination by outgassing.
The chamber design allowed the use of a maximum xenon gas pressure of 2 atm.

Neutrons scattered by the xenon gas target were measured by a multiwire  proportional chamber (MWPC) 
containing a mixture of 60\% ${}^3$He and 40\% CF${}_4$.
The active volume of the MWPC was 980 mm (horizontal) $\times$ 980 mm (vertical) $\times$ 63.5 mm (depth),
and its detection efficiency was 80\% for neutrons of wavelength 5 \AA.
The MWPC consisted of a single anode wire plane sandwiched by two cathode wire planes at a distance of 6.4 mm.
Each plane consisted of 192 wires at a pitch of 5.1 mm, giving a spatial resolution of 5 mm (FWHM).
The front (rear) cathode plane was oriented vertically (horizontally),
so the detector can be treated as a two--dimensional pixelated detector with pixel size 5.1 $\times$ 5.1 mm$^2$.
Calibration of the relative detection efficiencies of each pixel to an accuracy of 0.5\% was performed using a
standard water sample \cite{Wignall:1987du}.
The detector was placed 3.11 m downstream of the center of the target chamber.
The entire neutron beam path was evacuated to 0.1 Pa to minimize scattering off residual gas.

%=== fig 3 =====
\begin{figure}[b]
\includegraphics[width=8.2cm]{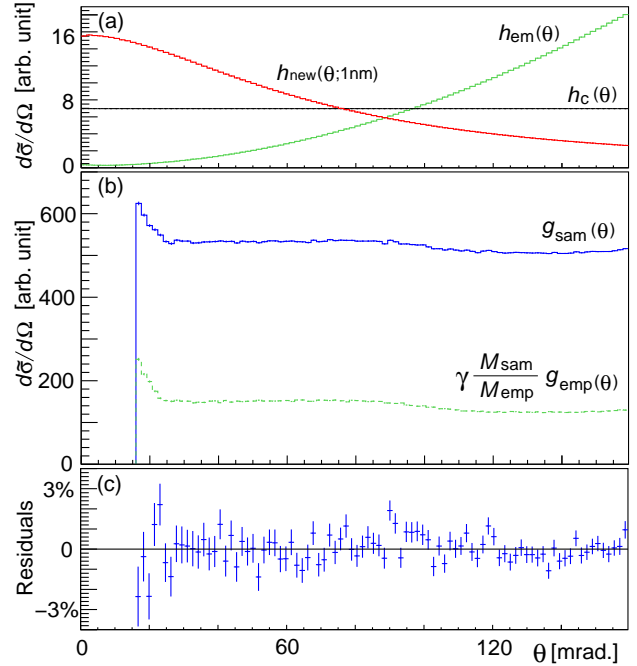}%
\caption{\label{fig3}
(a) 
Simulated corrected differential cross sections for the constant term $h_{c}(\theta)$,
the electromagnetic interaction term $h_{em}(\theta)$, and 
the new interaction term $h_{new}(\theta;\lambdabar=1$nm$)$.
They are normalized to $\int_{S} h(\theta)d\theta = 1$ over the signal acceptance $S$
of 16.1 mrad $ < \theta < $ 160.8 mrad.
Within this signal region, the distribution due to the new interaction term for $\lambdabar \sim 1$ nm
is clearly different from the known interaction terms.
(b)
Measured distributions for the xenon-filled [$g_{sam}(\theta)$]
and empty [$g_{emp}(\theta)$] chamber.
The contribution from beam background $g_{bg}(\theta)$, not shown in the figure,
has a flat distribution, with around 3 counts/bin.
(c) 
Residuals from the reference distribution due to known interactions.
The data, shown as crosses, are consistent with the case of no new forces
($\chi^2/$ndf$ \,\, = 1.4$).
}
\end{figure}
%=== fig 3 =====
The differential scattering cross section of unpolarized neutrons on xenon atoms
due to known and new interactions can be expressed as
\cite{Sears:1986go}
\begin{eqnarray}
\frac{d\sigma}{d\Omega} (q) & = & 
b_c^2 \left\{ 1 + \chi_{em}[1-f(q)] + \chi_{new} \frac{\mu^2}{q^2 + \mu^2} \right\}^2 \nonumber \\
&& + b_{s}^2(q) + b_i^2 + O(b_{F}^2)  \\
& \simeq & b_c^2 \left\{ 1 + 2\chi_{em}[1-f(q)] + 2\chi_{new} \frac{\mu^2}{q^2 + \mu^2} \right\}, \label{eq:cs}
\end{eqnarray}
where $q$ is the momentum transfer, 
$b_c$ ($\sim 5$ fm) is the coherent scattering length,
$b_s(q)$ ($\sim 10^{-3}$ fm) is the Schwinger scattering length, 
$b_i$ is the incoherent scattering length (negligibly small in the case of the xenon atom),
and $b_F$ ($\sim 10^{-3}$ fm) is the Foldy scattering length.
$\chi_{em}$ represents the strength of the known non-constant term in the coherent scattering processes
due to electromagnetic interactions between the neutron's spin or charge distribution and atomic fields.
It is defined as $\chi_{em} \equiv (b_{F} + b_{I})Z/b_c$, 
where $b_{I}$ ($\sim 10^{-3}$ fm) is the intrinsic neutron-electron scattering length and $Z$ is the atomic number.
$\chi_{em}$ is of order $10^{-2}$.
The atomic form factor can be described to an accuracy of $10^{-4}$
by the empirical formula $f(q) = [1+3(q/q_0)^2]^{-0.5}$,
where $q_0 = 6.86$ \AA$^{-1}$ \cite{Sears:1986go}.
The additional scattering length due to a new Yukawa-type scattering potential
is given in the third term, where $\chi_{new} \equiv m_n g^2 Q_1 Q_2 /2\pi b_c \mu^2$.
With $Q_1$, $Q_2$ equal to the neutron and xenon nuclear masses,
$\mu = 200$ eV and $g^2 = 10^{-16}$ GeV$^{-2}$ give $\chi_{new} = 2 \times 10^{-3}$.
Equation\,(\ref{eq:cs}) is obtained
by neglecting terms smaller than $10^{-4}$ fm$^2$.

The expected angular scattering distribution to be measured was derived
from this differential cross section convoluted with the finite beam size, 
the length of the scattering chamber, and the thermal motion of the xenon gas,
which follows the Maxwell-Boltzmann distribution at 293 K.
These effects were simulated using the Monte-Carlo method,
and a corrected differential cross section $d\tilde{\sigma}/d\Omega(\theta)$ 
corresponding to this experimental setup was expressed as
the sum of three terms $h_{c}(\theta)$, $h_{em}(\theta)$, and $h_{new}(\theta; \mu)$,
corresponding to the constant, electromagnetic, and new interaction terms of Eq.\,(\ref{eq:cs}):
\begin{eqnarray}
\frac{d\tilde{\sigma}}{d\Omega}(\theta) = N && \left\{ (1 - \alpha^{*})(1 - \beta) h_{c}(\theta) \right.\nonumber \\
&& + \alpha^{*}(1 - \beta) h_{em}(\theta) + \left. \beta h_{new} (\theta; \mu) \right\}~. \label{eq:fit}
\end{eqnarray}
The distributions are normalized to
$\int_{S} d\tilde{\sigma}/d\Omega(\theta) d\theta = N$ and $\int_{S} h(\theta) d\theta = 1$
over the signal acceptance $S$ of 16.1 mrad $ < \theta < $ 160.8 mrad,
and $\alpha^{*}:1-\alpha^{*}$ is the ratio of scattering contributions due to the electromagnetic and constant terms.
$\alpha^{*}$ is determined to be $(1.09 \pm 0.01)\times 10^{-4} $
from measurements of the neutron mean-square charge radius $\left<r_{n}^{2}\right> = -0.1161 \pm 0.0022$ fm$^{2}$ \cite{PDG-2014}.
$\beta$ is the quantity to be estimated, the fraction due to new forces.
Figure\,\ref{fig3}(a) shows simulated distributions for these three terms.

The distribution of neutrons was measured
for beam backgrounds $g_{bg}(\theta)$ for 3 h,
with an empty target chamber $g_{emp}(\theta)$ for 36 h, 
and with a xenon-filled chamber $g_{sam}(\theta)$ for 72 h, separately.
The corresponding number of 
counts registered by the flux monitor are $M_{bg}$, $M_{emp}$, and $M_{sam}$.
The scattering contribution due to the xenon gas sample $g(\theta)$ was estimated as
\begin{eqnarray}
g(\theta) = g_{sam}(\theta) 
	&-& \gamma\frac{M_{sam}}{M_{emp}} g_{emp}(\theta)\nonumber \\
	&& - (1 - \gamma)\frac{M_{sam}}{M_{bg}} g_{bg}(\theta)~,
\end{eqnarray}
where $\gamma$ is the transmission of the xenon gas, 
measured to be $0.904 \pm 0.004$. 
Fig.\,\ref{fig3}(b) shows the distributions $g_{sam}(\theta)$ and $g_{emp}(\theta)$.
The net distribution from the xenon gas is well described by a linear sum of only the $h_{c}(\theta)$ and $h_{em}(\theta)$ terms
(corresponding to $\beta = 0$, or no new forces),
for which the $\chi^2$ per degree of freedom is $127/90 = 1.4$.
The residuals from the reference distribution due to known interactions are shown in Fig.\,\ref{fig3}(c).

%=== fig 4 =====
\begin{figure}
\includegraphics[width=8.2cm]{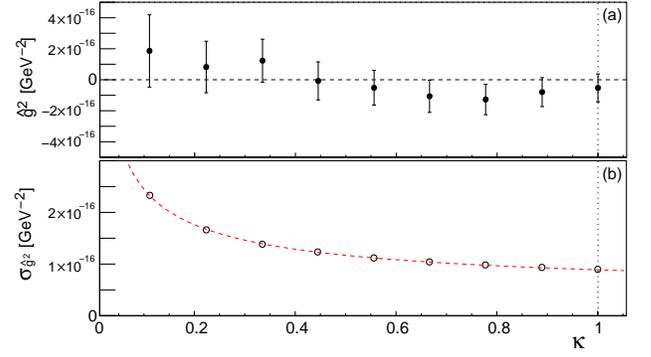}
\caption{\label{fig4}
(a) 
Evolution of the best fit values $\hat{g^{2}}$ as a function of the fraction of collected events $\kappa$
for $\lambdabar = 1$ nm.
The full data set corresponds to $\kappa = 1$.
(b)
Evolution of the estimated $\hat{g^{2}}$ uncertainty $\sigma_{\hat{g}^{2}} (\kappa)$.
It is well described by the simple model $\sigma_{\hat{g}^{2}} (\kappa) = (t/\sqrt{\kappa}) \oplus u$,
with $t = 7.6 \times 10^{-17}$ GeV$^{-2}$ and $u = 4.8 \times 10^{-17}$ GeV$^{-2}$,
shown as the dashed curve.
}
\end{figure}
%=== fig 4 =====
%=== fig 5 =====
\begin{figure}
\includegraphics[width=8.2cm]{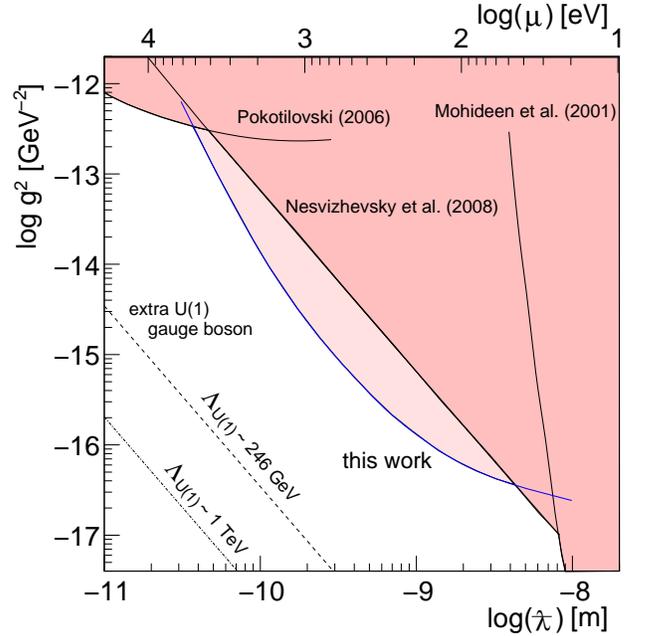}%
\caption{\label{fig5}
Comparison of the 95\% C.L. limits presented in this Letter with those given in 
\cite{Pokotilovski:2006gd, Nesvizhevsky:2008jz,BORDAG:2001tt}.
The results improve the previous constraints by a factor of up to 10
for interaction ranges between 4 and 0.04 nm.
The dashed line shows the theoretical prediction due to extra U(1) gauge bosons
with symmetry braking scales of $\Lambda_{U(1)} \sim 246$ GeV and $ \sim 1$ TeV \cite{Fayet:1993fu, Fayet:2001ve}.
}
\end{figure}
%=== fig 5 =====
The measured scattering distribution was fitted by Eq.\,(\ref{eq:fit})
for several hypotheses of a new boson's interaction range.
As examples, the best fit values of $\beta$ were 
$\hat{\beta} = (0.1 \pm 2.7) \times 10^{-2}$ at $\lambdabar = 0.1$ nm 
and 
$\hat{\beta} = (-0.7 \pm 1.2) \times 10^{-3}$ at $\lambdabar = 1.0$ nm, 
corresponding to
$\hat{g}^2 = (0.2 \pm 6.8) \times 10^{-15}$ GeV$^{-2}$ 
and $\hat{g}^2 = (-5.3 \pm 9.0) \times 10^{-17}$ GeV$^{-2}$, respectively.
The systematic effect on the extracted limits due to the uncertainty of the $\gamma$ measurement is estimated to be 
$\pm 15\%$($\lambdabar = 0.1$ nm) and $\pm 18\%$($\lambdabar = 1$ nm),
and less than $0.1\%$ due to the uncertainty on $\alpha^{*}$.
Additional effects due to the neutron wavelength determination, gas temperature, pixel efficiency calibration,
and atomic form factor modeling were tested using a pseudoexperiment technique
and confirmed to be significantly smaller than the obtained sensitivities.
To check for unexpected time-dependent systematic effects,
the evolution with the increasing data set size of $\hat{g}^2$ extracted for $\lambdabar = 1$ nm and its error $\sigma_{\hat{g}^2}(\kappa)$ 
are shown in Figs.\,\ref{fig4}(a) and \ref{fig4}(b).
The errors are well described by the simple model
$\sigma_{\hat{g}^2}(\kappa) = (t/\sqrt{\kappa}) \oplus u$,
where $\kappa$ is the data set fraction,
and $t$ and $u$ are constants, determined to be $t = 7.6 \times 10^{-17}$ GeV$^{-2}$ and $u = 4.8 \times 10^{-17}$ GeV$^{-2}$.
No additional sources of significant systematic uncertainties were identified.

Limits on $g^2$ at 95\% confidence level were evaluated using the Feldman and Cousins approach \cite{Feldman:342419}.
The obtained upper limit curve is shown in Fig.\,\ref{fig5}.
These results improve previous constraints for gravitylike forces in the 
4 to 0.04 nm range by a factor of up to 10.

\begin{acknowledgments}
% put your acknowledgments here.
We are grateful to Young-Soo Han, Tae-Hwan Kim, and Eun-Hye Kim 
for the stable operation of the neutron beam line and for significant engineering support.
We thank Toshinori Mori and Wataru Ootani for their valuable suggestions on xenon gas operations,
and thank Daniel Jeans for useful discussions about statistical treatment.
This work is supported by JSPS KAKENHI Grant No. 90434323.
\end{acknowledgments}

% Create the reference section using BibTeX:
\bibliography{SansGrav_PRL}

\end{document}